\documentclass[prd,eqsecnum,superscriptaddress,preprint,tightenlines]{revtex4}
 \usepackage{amsmath} 
\usepackage{graphicx}
\usepackage{mathtools}

\DeclarePairedDelimiter{\ev}{\langle}{\rangle}
\DeclarePairedDelimiter{\abs}{\lvert}{\rvert}%
\newcommand{\Ai}{\operatorname{Ai}}
\newcommand{\Bi}{\operatorname{Bi}}

\newcommand{\DD}[2]{\frac{\partial^2 #1}{\partial #2^2}}
\newcommand{\PCyl}[2]{\operatorname{D}_{#1}(#2)}
\renewcommand{\vec}[1]{\text{\boldmath$#1$}}

\begin{document}
 
 \title{Vacuum energy density and pressure near  a soft wall}
 
  \author{S. W.  Murray}
 \email{borgbilly@tamu.edu}
\affiliation{Department of Physics and Astronomy, Texas A\&M 
University,   College Station, TX 77843-4242, USA}

 \author{C. M. Whisler}
 \email{whisler2@wisc.edu} 
\affiliation{Department of Physics and Astronomy, Texas A\&M 
University,   College Station, TX 77843-4242, USA}
\affiliation{Department  of Mathematics,  Texas A\&M 
University, College Station, TX 77843-3368, USA}
\affiliation{
Present address:  Department of Physics, University 
of Wisconsin, 1150 University Ave., Madison, WI 53706, USA}

\author{S. A. Fulling}
\email{fulling@math.tamu.edu}
 \homepage{http://www.math.tamu.edu/~fulling}
\affiliation{Department of Physics and Astronomy, Texas A\&M 
University,   College Station, TX 77843-4242, USA}
\affiliation{Department of Mathematics,  Texas A\&M 
University, College Station, TX 77843-3368, USA}

\author{Jef Wagner}
\email{jef.wagner@lawrence.edu}
\affiliation{Department of Mathematics, Texas 
A\&M University,   College Station, TX 77843-3368, USA}
\affiliation{Present address: Department of Physics,  Lawrence 
University, Appleton, WI 54911,  USA}

\author{H.~B. Carter}
\email{hcarter333@email.tamu.edu}
\affiliation{Department of Physics and Astronomy, Texas A\&M 
University,   College Station, TX 77843-4242, USA}

\author{David Lujan}
\email{dlujan-94@tamu.edu}
\affiliation{Department of Physics and Astronomy, Texas A\&M 
University,   College Station, TX 77843-4242, USA}

\author{F. D. Mera}
\email{merandi12@gmail.com}
\affiliation{Department of Mathematics, Texas A\&M 
University,   College Station, TX 77843-3368, USA}

\author{T. E. Settlemyre}
\email{tommy7410@tamu.edu}
\affiliation{Department of Physics and Astronomy, Texas A\&M 
University,   College Station, TX 77843-4242, USA}
\affiliation{Department of Mathematics,  Texas A\&M 
University, College Station, TX 77843-3368, USA}

 \date{March 16, 2016}

\begin{abstract}
Perfectly conducting boundaries, and their Dirichlet 
counterparts for quantum scalar fields, predict 
nonintegrable energy densities.  A more realistic
model with a finite ultraviolet cutoff yields two
inconsistent values for the force on a curved or edged
 boundary (the ``pressure anomaly'').  A still more
 realistic, but still easily calculable, model replaces
 the hard wall by a power-law potential; because it
 involves no \emph{a posteriori} modification of the formulas 
calculated from the theory,
  this model should  be anomaly-free. 
 Here we first set up the formalism and
 notation for the quantization of a scalar field in
 the background of a planar soft wall, and we 
 approximate the reduced Green function in perturbative and
 WKB limits (the latter being appropriate when either
 the mode frequency or the depth into the wall is 
 sufficiently large).  Then we display numerical 
 calculations of energy density and pressure
 for the region outside the wall, which
 show that the pressure anomaly does not occur there.
 Calculations inside the wall are postponed to later papers,
 which must tackle regularization and renormalization 
 of divergences induced by the potential in the bulk
 region.

  \end{abstract} 
 
 \maketitle

  \section{Introduction} \label{sec:intro}

The Casimir effect \cite{Milton,BKMM,Dalvit} is 
traditionally thought of as a force between electrical
 conductors
caused by the effects of these conducting boundaries 
on the spectrum of normal modes of the quantized 
electromagnetic field
in the region exterior to the conductors.
(For simplicity when investigating matters of principle,
the EM field is often replaced by a scalar field, and the
boundary conditions of a perfect conductor replaced by
the Dirichlet condition, $\phi=0$. The present work deals
with that model.)
Crudely, the idea is that each mode acts as a harmonic 
oscillator with ground-state energy $\frac{1}{2} \omega$,
so that when the energy is summed over all modes of the system,
 the total vacuum energy is $\ev{E} = \frac{1}{2} \sum_{n} 
\omega_n\,$; then the derivative of $\ev{E}$ with respect to
some geometrical parameter of the system constitutes a 
generalized force, by the 
``principle of virtual work''\negthinspace.

This mode sum  is clearly divergent, and its relation to 
physical realities 
measured in the laboratory has been somewhat controversial.
Formal regularizations, 
such as the zeta-function method \cite{Elizalde}, 
usually give answers that are generally accepted as 
``correct''\negthinspace, but their physical logic is
unconvincing.
Also, when the conductors are regarded as rigid and only
their relative positions are allowed to vary, a finite answer
for the force (or energy difference) can be obtained by
sufficiently careful calculations \cite{BartonA,Jaffe,Schaden}.
It is generally agreed, however, that the root of the problem
is the failure of the idealized notion of a perfect conductor.
A real material does not act like a perfect conductor 
(or even a perfect dielectric) at
very high frequencies, and thus nature must provide an
ultraviolet (or high-momentum \cite{BartonB}) cutoff.
But the detailed study of a real material is a complicated
nonlinear problem of condensed-matter physics, 
and it is of interest to step back and see if the phenomena
can still be modeled, at least qualitatively, 
entirely within the linear framework
of the field alone.

The most common approach is to insert \emph{ad hoc}
an ultraviolet cutoff to make the integrals over frequency 
finite. 
The simplest choice is an exponential cutoff:
$$\ev{E}_\tau \equiv
\frac{1}{2} \sum_n \omega_n e^{-\omega_n \tau},$$
 where $\tau$ is the cutoff parameter, which can also be 
considered as a  Wick rotation of the difference of two time 
coordinates, 
$\tau = -i (t - t')$. 
A complementary approach is to calculate expectation values
of  the local energy density and pressure
(components of the stress tensor:  
$u\equiv \ev{T_{00}}\,$, $p_x \equiv \ev{T_{11}}\,$, etc.), 
 from which all the physically interesting quantities are
 in principle  derivable.
 From local calculations one sees that, once one removes
 the ubiquitous zero-point energy that is present even for
 a completely free field, the vacuum energy density at any
 point away from the boundary is finite even without a cutoff,
 and infinities in the total energy (for a region of finite
 volume) arise only because the energy density is not
 integrable near the boundary.
 Thus this approach significantly elucidates the physical 
 meaning of the divergences.
  Cutoff and local calculations fit very well together, as
  both can be based on a Green function of the field equation 
corresponding to the geometry of the system considered 
(e.g., \cite{rect, HSW}). 

It was reasonable to expect that keeping $\tau$ nonzero, and
comparable (in natural units) to an interatomic spacing,
would yield an effective model of the Casimir effects of
 a realistic imperfect conductor.
However, when ultraviolet regularization is applied to
the stress tensor, it produces formulas for energy 
density and pressure that are inconsistent with each other
\cite{norman09, Estrada-Mera}. 
 The principle of virtual work, or energy-pressure balance,
  requires that a change in the vacuum energy corresponding
   to some 
infinitesimal movement of a boundary of the system 
be attributable 
to some vacuum pressure pushing against that boundary. 
As an example, consider the pressure against a flat wall,
\begin{equation}\label{balance}
  -\,\frac{\partial E}{\partial x} = F = \iint p\, dy\, dz
  \qquad \left(E = \iiint u\, dx\,dy\,dz\right).
\end{equation}
This relation does not follow by default from the local 
energy-momentum conservation law, 
${\partial T^{\mu \nu}}/{\partial x^\mu}=0$
(which is satisfied by the expectation values inside the
 cavity).
It instead constrains the equation of state of the 
quantized field (cf.~\cite{TAnomaly}). 
It appears that the principle of virtual work has been
 disrupted by the regularization method.
 Pressure balance  can be restored by replacing
  the timelike ultraviolet cutoff by
point-splitting in a ``neutral'' direction,
one parallel to the wall 
\cite{Estrada-Mera,Benasque}, but this is to repair 
the damage done by one
\emph{ad hoc} measure by adding another one.

Therefore, Bouas et al.~\cite{SpecGeom} proposed a new
kind of model, the \emph{soft wall}, wherein 
 the Dirichlet boundary is replaced by a smooth, steeply 
rising potential function.
Although regularization is still needed, as always in  quantum
field theory, the pathology of the perfect conductor has
been removed in a physically rigorous way.
That is, the departure from ideal conductivity is embodied
from the start in a complete and consistent physical theory, 
rather than imposed \emph{a posteriori} by inserting a cutoff
into the formal mathematical results of an unrealistic theory.
One has every reason to expect the principle of virtual work
to hold (after renormalization) in this context.

Specifically, \cite{SpecGeom} and \cite{HSW}, summarized in
 \cite{Benasque}, studied a  ``power wall''
 set at the plane $z=0$ (see Eq.~(\ref{potential})).
It models a conducting plate whose thickness is much greater 
than its skin depth and whose size is 
large enough that edge effects are negligible.
That project was continued in the thesis \cite{thesis},
whose results are reported here with some corrections.   
Following Milton \cite{HSW},  we express the components of the 
vacuum stress-energy tensor in terms of the 
\emph{reduced Green function} formed from the 
solutions of the $z$ dependence
 of the field in the imaginary-frequency regime (Sec.~II).
 In Sec.~III we study those solutions analytically (in the
 limits of large and small frequency) and numerically.
 Energy density and pressure in the region outside
 the potential are plotted in Sec.~IV;
 there, as expected,
  we find no indication of an anomaly in the
 energy-pressure balance.
Sec.~V sets up some machinery to be used in the future to
tackle the more difficult problem of the interior of the wall,
and Sec.~VI summarizes the results and the outlook.
 Some references to older or tangentially related work that did 
not fit into this introduction are provided in an appendix.

\section{Formulation of the model} \label{sec:model}

\subsection{Field equation, basis solutions, and Green function} 
 \label{sec:field}

 We consider the scalar field equation
\begin{equation}\label{fieldeq}
        \DD{\Phi}{t} = \nabla^2 \Phi - V(z) \Phi,
\end{equation}
where 
\begin{equation} \label{potential}
        V(z) =
        \begin{cases}
                0 & \text{if } z \leq 0, \\
                z^{\alpha} & \text{if } z > 0,
        \end{cases}
\end{equation}
is a scalar potential (or position-dependent Klein--Gordon mass).
The field is classically real, hence Hermitian as a quantum 
field.
The space-time coordinates are $(t,x,y,z) = (x^0,x^1,x^2,x^3) = 
\underline x $,
and the metric tensor has signature $(-1,1,1,1)$.
One often writes $\vec{r_{\perp}}$ for $(x,y)$,
and $\vec{k_\perp}$ for the conjugate wave numbers, $(k_1,k_2)$.
We use the natural units $\hbar =1 =c$ and in addition choose the 
unit of length so that the coupling constant implied in 
Eq.~(\ref{potential}) is unity;
the fundamental length scale that emerges for a general coupling 
constant is identified in \cite{SpecGeom}.

To avoid introducing gratuitous singular behavior around $z=0$,
we  take $\alpha$ to be a positive integer.  As 
$\alpha\to\infty$,  the potential formally approaches 
\begin{equation}
        V(z) =
        \begin{cases}
                0 & \text{if } z < 1, \\
                \infty & \text{if } z > 1,
        \end{cases}
\end{equation}
which represents a hard wall (Dirichlet boundary condition)
 at $z = 1$.
Our goal is to understand the ground-state expectation values of 
energy density and pressure, which will be of significant 
magnitude only in some interval surrounding the interval 
$0\le z\le1$.

When Eq.~(\ref{fieldeq}) is solved by separation of variables,
the dimensions $(t,x,y)$ are trivial, and the $z$ dependence
is given by solutions of
\begin{equation}\label{eigeneq}
\bigg(- \frac{\partial^2}{\partial z^2} + V(z) -
p^2 \bigg) \phi (z) = 0,
\end{equation}
as detailed in \cite{SpecGeom}. 
However, following \cite{HSW}, we find it more useful to 
express vacuum expectation values in terms of a reduced Green
function, built out of solutions of Eq,~(\ref{eigeneq}) with the 
sign of the spectral parameter reversed:
\begin{equation}\label{redgreeneq}
\bigg(- \frac{\partial^2}{\partial z^2} + V(z) +
\kappa^2 \bigg) \phi_{\kappa} (z) = 0.
\end{equation}

For each (positive) value of $p$ or $\kappa$, there are two 
linearly independent solutions of Eq.\ (\ref{eigeneq}) or 
(\ref{redgreeneq}). Most obviously relevant are the  solutions 
that decay as $z\to +\infty$;  
let $F(z)$ be such a solution, normalized 
at the origin for definiteness:
\begin{equation} \label{Fdef}
        F(0) = 1,  \qquad \lim_{z \to \infty} F(z) = 0.
\end{equation}
A convenient choice of second solution is defined by
\begin{equation}\label{Gdef}
        G(0) = 0,\qquad G'(0) = 1.
\end{equation}
In constructing the Green function, however, the most pertinent 
second solution is one that decays at $-\infty$:
\begin{equation}\label{Hdef}
H(0)=1, \qquad \lim_{z \to -\infty} H(z) = 0.
\end{equation}

It is sometimes convenient to relax the normalization conventions
$F(0)=G'(0)=H(0)=1$, and even the condition $G(0)=0$ (provided 
that $G$ and $F$ stay independent).
When we  give formulas that do not make those assumptions, we 
write the function names in the calligraphic font.
For example, one
notes immediately that for $z\le 0$,
\begin{equation}\label{Hneg}
H_\kappa(z)=  e^{\kappa z},
\end{equation}
and hence that
\begin{equation}\label{Hderiv}
\kappa = H_\kappa'(0)
 = \frac{{\cal H}_\kappa'(0)}{{\cal H}_\kappa(0)}\,.
\end{equation}

For $z\ge0$,
the case $\alpha = 1$ is easily solved using Airy functions, and 
the case 
$\alpha = 2$ also produces an exact solution involving parabolic 
cylinder functions.
 For $\alpha = 1$, the solutions  (normalized as above) are
\begin{align} 
        F_\kappa(z) &= \frac{\Ai(\kappa^2 + z)}{\Ai(\kappa^2)} 
\,,
\label{airyF}\\
        G_\kappa(z) &= \frac{\Bi(\kappa^2) \Ai(\kappa^2 + z) - 
\Ai(\kappa^2) \Bi(\kappa^2 + z)}{\Ai '(\kappa^2) \Bi(\kappa^2) 
- \Ai(\kappa^2) \Bi' (\kappa^2)}\,. \label{airyG}
\end{align}
For $\alpha = 2$, the solutions are
\begin{align}
        F_\kappa(z) &= \frac{\PCyl{-(\kappa^2 + 
1)/2}{\sqrt{2}z}}{\PCyl{-(\kappa^2 + 1
)/2}{0}}\,, \label{parcylF} \\
        G_\kappa(z) &= \frac{\PCyl{-(\kappa^2 + 
1)/2}{-\sqrt{2}z}-\PCyl{-(\kappa^2 + 1
)/2}{\sqrt{2}z}}{2\sqrt{2} \PCyl{(1-\kappa^2)/2}{0}}\,.
\label{parcylG}\end{align}
\begin{figure}
        \centering
                \includegraphics[scale=.5]{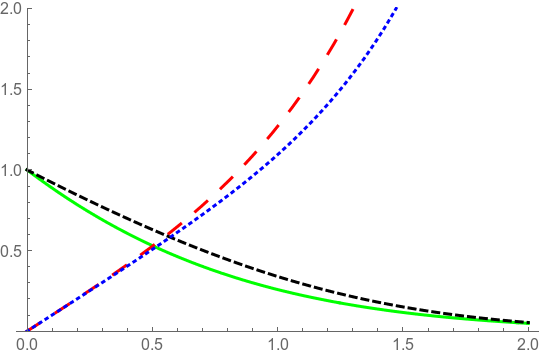}
 \caption{(color online) $F_\kappa(z)$ (solid, green) and 
$G_\kappa(z)$ 
(long dashed, red) for $\alpha = 1$, $\kappa = 1$,
and $F_\kappa(z)$ (dashed, black) and 
$G_\kappa(z)$ (dotted, blue) for $\alpha = 2$, $\kappa = 0.5$,
The differences between the two pairs of curves (in this region 
of small~$z$) are caused more by the change in $\kappa$ than by 
the change in $\alpha$. All graphics in this paper were prepared 
with {\sl Mathematica}.} 
 \label{swo1}\end{figure}
In reference to the denominator in Eq.~(\ref{parcylG}) see 
\cite[(19.6.2) and (19.3.5)]{NBS}.
Examples of these functions are plotted in
Fig.~\ref{swo1}.
Unfortunately, there are no known closed-form solutions to the 
differential equation when $\alpha > 2$. 
Since we intend to examine the limiting 
case as $\alpha \to \infty$, we will need to find approximate 
solutions to these equations. 

We denote the Wronskian of two solutions by
\begin{equation}\label{wronskian}
W(y_1,y_2) = y_1 y_2' - y_1' y_2\,,
\end{equation}
where the functions may be evaluated at any value of $z$, in 
particular at $z=0$.

The reduced  Green function is to satisfy
\begin{equation} \label{greeneq}
\bigg(- \frac{\partial^2}{\partial z^2} + V(z) + \kappa^2 \bigg) 
g_{\kappa} (z,z') = \delta(z-z').
\end{equation}
By a well known formula or method, one finds
\begin{equation}\label{greenH}
g_\kappa(z,z') = \frac{{\cal H}_\kappa(z_<) {\cal F}_\kappa(z_>)}
{W({\cal F}_\kappa,{\cal H}_\kappa)}
=\frac{{ H}_\kappa(z_<) {F}_\kappa(z_>)}
{\kappa-F'_\kappa(0)}
\,,\end{equation}
\begin{equation}\label{zpm}
z_<= \min(z,z'), \qquad z_>=\max(z,z').
\end{equation}
Further reduction is postponed to later sections, because
the most useful approach depends on whether one is working 
outside or inside the wall.

\subsection{Stress tensor} \label{sec:stress}

The  stress-energy-momentum tensor of a massless 
scalar field in flat space-time  with curvature-coupling 
(conformal) parameter $\xi$ is 
\begin{align}
T_{\mu \nu} =& \partial_\mu \Phi \partial_\nu \Phi - \frac{1}{2} 
g_{\mu \nu} (\partial_\lambda \Phi \partial^\lambda \Phi + V \Phi^2)
 \nonumber \\ 
&- \xi (\partial_\mu \partial_\nu - g_{\mu \nu} \partial_\lambda 
\partial^\lambda) \Phi^2,
\label{Tmunu}\end{align}
where $g_{\mu \nu} =
\eta_{\mu \nu} = \text{diag} (-1,1,1,1)$.
The equation of motion (\ref{fieldeq}) 
can be used  \cite{Gravity}
to rewrite Eq.~(\ref{Tmunu}) so that $V$ does not explicitly 
appear. Also, the notation
\begin{equation}\label{beta}
\beta = \xi -  \tfrac{1}{4}
\end{equation}
is convenient.
Thus one arrives at these formulas for 
 the energy density and pressure:
\begin{subequations} \label{energypressure}
\begin{align} 
T_{00} &= \frac{1}{2} (\partial_0 \Phi)^2 - \frac{1}{2} \Phi 
(\partial_0)^2 \Phi - \beta \nabla^2 \Phi^2, \label{energyden}\\
T_{11} &= \frac{1}{2} (\partial_1 \Phi)^2 - \frac{1}{2} \Phi 
(\partial_1)^2 \Phi + \beta (-\partial_0^2 + \partial_2^2 + \partial_3^2)
 \Phi^2, \label{1pressure}
\end{align}
\end{subequations}
and totally analogous expressions for  
 $T_{22}$ and $T_{33}\,$.
 The remaining, off-diagonal, components of 
 $ T_{\mu \nu}$ 
play almost no role in our study;
it is easy to show that their vacuum expectation values must
vanish for a single, flat boundary as considered here 
(although  they may temporarily develop nontrivial 
cutoff-dependent terms when point-splitting regularization is 
used with 
an oblique direction of point separation \cite{christensen}).

The expression of the vacuum expectation values
 $ \langle T_{\mu \nu}\rangle$ in terms of Green functions has 
been detailed in previous papers \cite{SpecGeom,HSW,Benasque}, 
but in a variety of formalisms.  Here we shall not rehearse the 
standard quantum field theory leading to the main results, but
rather concentrate on establishing consistent notations.
On the one hand, the expectation value of the product of two 
field operators is a certain Green function for the wave 
equation,
\begin{equation}\label{vev}
\ev{0|\Phi(\underline x) \Phi(\underline x')|0} = \frac{1}{i} 
G(\underline x,\underline x') ,
\end{equation}
of the form
\begin{equation}\label{wightman}
G(\underline x,\underline x') = \int \frac{d \omega}{2 \pi} 
\frac{d \vec{k}_\perp}{(2 \pi)^2} 
e^{-i \omega (t-t')} e^{i \vec{k}_\perp \cdot (\vec{r}_\perp 
-\vec{r}'_\perp)} 
g(p^2;z,z')
\end{equation}
(all three integrations being over $(-\infty,\infty)$), with
\begin{equation} \label{kappawave}
p^2= \omega^2 - \vec{k}_\perp^2
\end{equation}
(which, despite the notation, is not necessarily positive).
 The vacuum energy density and pressure can now be found 
from Eq.~(\ref{wightman}) and the expectation values of
Eqs.\ (\ref{energypressure}).
If we set
\begin{equation} \label{wick}
\omega = i \zeta, \qquad t-t' = i (\tau -\tau'),
\end{equation}
and formally rotate the $\zeta$ integration back to the real 
axis, we get
\begin{equation}\label{Gcyl}
-iG(\underline x,\underline x') = \int \frac{d \zeta}{2 \pi} 
\frac{d \vec{k}_\perp}{(2\pi)^2} e^{i (\zeta \tau + 
\vec{k}_\perp \cdot \vec{r}_\perp)} 
g_{\kappa}(z,z'),
\end{equation}
which we have here simplified by setting 
$\tau'$
and $\vec{r}'_\perp$ equal to zero.
We now have
\begin{equation} \label{kappacyl}
\kappa^2= \zeta^2 + \vec{k}_\perp^2,
\end{equation}
which is always positive.
On the other hand, the calculation of the expectation values can 
be based from the outset on the ``cylinder kernel''\negthinspace,
a certain Green function for the equation
\begin{equation}\label{cyleq}
        \DD{\Phi}{\tau} + \nabla^2 \Phi - V(z) \Phi = 0.
\end{equation}
Standard construction of that function leads rigorously to 
Eq.~(\ref{Gcyl}), times $-2$:
\begin{equation}\label{Tcyl}
\overline{T}(\tau,\vec{r}'_\perp, z,z') = -2 \int \frac{d 
\zeta}{2 \pi} 
\frac{d \vec{k}_\perp}{(2\pi)^2} e^{i (\zeta \tau + 
\vec{k}_\perp \cdot \vec{r}_\perp)} 
g_{\kappa}(z,z'),
\end{equation}
with $\kappa$ defined as the positive root of Eq.~(\ref{kappacyl}).

 The next step of the calculation is most simply expressed by 
writing the vacuum stress as a Fourier transform:
\begin{equation} \label{tmunu}
\ev{T_{\mu \nu}} 
= \int \frac{d \zeta}{2 \pi} \frac{d \vec{k}_\perp}{(2 \pi)^2} 
e^{i \zeta \tau} e^{i \vec{k}_\perp \cdot \vec{r}_\perp} t_{\mu \nu}
 \Big|_{z' \rightarrow z}.
\end{equation}
From Eqs.\ (\ref{energypressure}), (\ref{vev}),  and 
(\ref{Gcyl}) one gets
\begin{subequations} \label{GtoT}
\begin{align}
t_{00} &= -\zeta^2 g_{\kappa}(z,z) - 2 \beta (\partial_z^2 + 
\partial_z \partial_z') g_{\kappa}(z,z') \big|_{z' \rightarrow 
z},
\label{Gtou} \\
t_{11} &= k_1^2 g_{\kappa}(z,z) + 2 \beta (\partial_z^2 + 
\partial_z \partial_z') g_{\kappa}(z,z') \big|_{z' \rightarrow 
z},
\label{Gtopx}\\
t_{22} &= k_2^2 g_{\kappa}(z,z) + 2 \beta (\partial_z^2 + 
\partial_z \partial_z') g_{\kappa}(z,z') \big|_{z' \rightarrow 
z},
\label{Gtopy} \\
t_{33} &=-\, \frac{1}{2} (\partial_z^2 - \partial_z \partial_z') 
g_{\kappa}(z,z') \big|_{z' \rightarrow z}. \label{gtopz}
\end{align}
\end{subequations}
It is noteworthy that $\ev{T_{33}}$ is independent of $\beta$
(or $\xi$).

Note that  primed coordinates other than $z'$, 
and derivatives with respect to them,  have been 
eliminated, so without loss of 
generality we may set those coordinates equal to~$0$.
Now the 4-vector 
\begin{equation}\label{deltavector}
\vec{\delta} = (\tau, \vec{r}_\perp, (z-z'))
\end{equation}
should be regarded as the separation between the two space-time 
points that are arguments of the fields in the quadratic field
observables; it is still needed to 
 regularize the formulas for the vacuum energy and pressure.
In the present work, however, we have no need to consider 
splitting in the $z$ direction, so we can set $z'=z$ as soon as 
the derivatives indicated in Eqs.\ (\ref{GtoT}) have been taken.
To understand the pressure anomaly it is essential, 
though, to consider point-splitting in an arbitrary direction in 
the $(\tau,\vec{r}_\perp)$ 3-space \cite{Estrada-Mera, sector1}.

Putting   Eqs.\ (\ref{tmunu}) and (\ref{GtoT}) together, we 
obtain formulas for $\langle T_{\mu\nu}\rangle$.
For example, the energy formula is
\begin{align}
\ev{T_{00}} =& \int \frac{d \zeta}{2 \pi} \frac{d 
\vec{k}_\perp}{(2 \pi)^2} e^{i \zeta \tau +i \vec{k}_\perp 
\cdot \vec{r}_\perp} \label{T00}\\
& \times [-\zeta^2 g_{\kappa}(z,z) - 2 \beta (\partial_z^2 + 
\partial_z \partial_z') g_{\kappa}(z,z') \big|_{z' \rightarrow z}]. 
\nonumber
\end{align}
After $z'$ is taken to $z$, the $\vec\delta$ of 
Eq.~(\ref{deltavector}) can  be thought of as a 3-vector and
the exponent in Eq.\ \ref{T00} can be written
$\vec{\kappa}\cdot\vec{\delta}$, where
\begin{equation}\label{kappavector}
\vec{\kappa} = (\zeta, \vec{k}_\perp).
\end{equation}
As in \cite{HSW}, we make a change of 
variables to polar coordinates, defined by
\begin{subequations}\label{polarcoords}
\begin{align}
\kappa^2 &= \abs{\vec{k}_\perp}^2 + \zeta^2, &\quad \cos{\theta} 
&= \frac{\zeta}{\kappa}\,, \\
\delta^2 &= \abs{\vec{r}_\perp}^2 + \tau^2, &\quad \cos{\phi} &= 
\frac{k_1}{\abs{\vec{k}_\perp}}\,. 
\end{align}
\end{subequations}
Note next that $z'$ can be completely eliminated from Eqs. 
(\ref{Gtou})--(\ref{Gtopy}) by recognizing the  total derivative
\begin{align}
\frac{d^2}{dz^2} & \Big[g_{\kappa}(z,z') \big|_{z' \rightarrow z} 
\Big] \nonumber \\ 
& = 2 \Big[\frac{\partial^2}{\partial z^2} g_{\kappa}(z,z') 
\big|_{z' \rightarrow z} +
 \frac{\partial}{\partial z} \frac{\partial}{\partial z'} 
g_{\kappa}(z,z') \big|_{z' \rightarrow z} \Big].
\end{align}
We then rewrite Eq.~(\ref{T00}) as
\begin{align}
\ev{T_{00}} =& \frac{1}{(2 \pi)^3} \Bigg( 
\frac{\partial^2}{\partial \tau^2} - 
\beta \frac{\partial^2}{\partial z^2} \Bigg) \int_{0}^{\infty} d 
\kappa\, \kappa^2 
\int_{-1}^{1} d \cos{\theta} \nonumber \\
& \times \int_{0}^{2 \pi} d \phi\, e^{i \kappa \sin{\theta} 
(\cos{\phi}\, r_1 + \sin{\phi}\, r_2)} e^{i \kappa \cos{\theta} 
\,\tau} 
g_{\kappa}(z,z)
\end{align}
\noindent and integrate over the angular coordinates. Doing the 
same for the other components, we finally arrive at the expectation
 values of the stress-energy tensor components:
\begin{subequations}\label{uandp}
\begin{align}
u\equiv \ev{T_{00}} &= \frac{1}{2 \pi^2} 
\bigg(\frac{\partial^2}{\partial 
\tau^2} - \beta \frac{\partial^2}{\partial z^2} \bigg) \int_{0}^{\infty} 
d \kappa\, \kappa g_{
\kappa}(z,z) \frac{\sin{\kappa \delta}}{\delta}\,, \label{u}\\
p_x\equiv\ev{T_{11}} &= \frac{1}{2 \pi^2} 
\bigg(-\frac{\partial^2}{\partial r_1^2} + \beta \frac{\partial^2}
{\partial z^2} \bigg) \int_{0}^{\infty} d 
\kappa\, \kappa g_{\kappa}(z,z) \frac{\sin{\kappa 
\delta}}{\delta} \,,\\
p_y\equiv\ev{T_{22}} &= \frac{1}{2 \pi^2} 
\bigg(-\frac{\partial^2}{\partial r_2^2} + \beta 
\frac{\partial^2}{\partial z^2} \bigg) \int_{0}^{\infty} d 
\kappa\,
\kappa g_{\kappa}(z,z) \frac{\sin{\kappa \delta}}{\delta}\,, \\
p_z\equiv\ev{T_{33}} &=-\, \frac{1}{4 \pi^2} \int_{0}^{\infty} d 
\kappa\, \kappa \bigg[\bigg(
 \frac{\partial^2}{\partial z^2} - \frac{\partial}{\partial z} 
\frac{\partial}{\partial z'} \bigg) g_{\kappa}(z,z') \bigg]
\bigg|_{z' \rightarrow z} \nonumber 
\\
& \quad \times \frac{\sin{\kappa \delta}}{\delta}\,.
\label{uandpz}\end{align}
\end{subequations}

 It is readily seen that the only dependence remaining 
on the point-splitting vector inside the integral is in the 
scalar $\delta$,
which depends on the differentiation variables 
$(\tau,\vec{r}_\perp)$ in a symmetrical way.
If there were no divergences, we would now find the 
physical energy density and pressure by taking
$\delta \rightarrow 0$.
 In that limit, formally 
 $u = -p_x = -p_y\,$, 
which is
 the expected relation between energy density and pressure as 
predicted by the principle of virtual work. 
To visualize this fact, place a test wall (say the $x$--$z$ 
plane)
perpendicular to the existing soft wall. If a pressure  $p_x$
pushes the test wall, there should be a decrease in energy 
corresponding to the amount of work done in the 
process of moving the test wall;
since $u$ is independent of the position of the test wall,
 this energy change is simply $u$ times the displacement.
Thus Eq.~(\ref{balance}) is satisfied --- in fact, the integrands 
on the two sides are pointwise the same.
Thus, \emph{insofar as the integrals in Eqs.\ (\ref{uandp}) 
converge},
 no pressure anomaly 
arises.
 We must now examine the 
precise situation outside and inside the wall separately
(Secs.\ \ref{sec:outside} and \ref{sec:inside}).

\section{Perturbation theory} \label{sec:pert}

When either $z$ or $\kappa$ is large, solutions of 
Eq.~(\ref{redgreeneq})  in the potential region 
are well approximated by the
(Carlini--Liouville--Green--Jeffreys--) WKB method for the 
exponential (nonoscillatory) regime.
Since the theory and formulas are well known, we shall simply 
introduce the  expressions when needed.

When both $z$ and  $\kappa$ are small, a different approximation 
method is needed.
One can write a solution of  Eq.~(\ref{redgreeneq})
 as a power series in~$\kappa^2$. 
To first order,
\begin{subequations}\label{FGpert}
\begin{align}
        F_\kappa(z) &\approx F_0(z) + \kappa^2 F_1(z), \\
        G_\kappa(z) &\approx G_0(z) + \kappa^2 G_1(z).
\end{align}
\end{subequations}
Luckily, the equation can be solved exactly when 
$\kappa = 0$. We introduce the  notations
\begin{equation}
        b = \frac{1}{\alpha + 2}\,, 
\end{equation}
\begin{subequations}\begin{align}
        k(z) &= \sqrt{z} K_{b}\! \left(2 b z^\frac{1}{2 
b} \right)\,, \\
        i(z) &= \sqrt{z} I_{b}\! \left(2 b z^\frac{1}{2 
b} \right)\,,
\end{align}\end{subequations}
where $K$ and $I$ are the modified Bessel  functions.
Then
\begin{subequations}
\begin{align}
F_0(z) &= c_1\, k(z), \\
G_0(z) &= c_2\, i(z),
\end{align}
\end{subequations}
where $c_1$ and $c_2$ are constants used to match the 
normalization conditions (\ref{Fdef}) and (\ref{Gdef}).

$F_1$  must be of the form
\begin{equation} \label{F1}
F_1(z)=\int_0^\infty g(z,z') F_0(z')\,dz',
\end{equation}
$g$ being  a suitable Green function for the nonhomogeneous
equation
\begin{equation} \label{nonhom}
F_1''-z^\alpha F_1=F_0\,.
\end{equation}
By hypothesis $c_1$ has already been chosen to match 
the boundary data exactly, so the proper solution 
must satisfy $F_1(0)=0$ as well as vanishing at infinity.
Therefore,
\begin{equation}\label{Fgreen}
g(z,z')=\frac{i(z_<)k(z_>)}{W(i,k)}\,.
\end{equation}
The solution for $G_1$ is similar, but this time the
Green function must annihilate both $G(0)$ and $G'(0)$,
with no condition at infinity.
Thus we arrive at
\begin{subequations}\label{pert1}\begin{align}
F_1(z) &=  \frac{1}{W(i,k)} \left(k(z) \int_{0}^{z} i(a) F_0(a) 
\,da + i(z) \int_
{z}^{\infty}k(a) F_0(a)\, da \right) ,\\
G_1(z) &= \frac{1}{W(i,k)} \left(k(z) \int_{0}^{z} i(a) G_0(a)\,da 
- i(z) \int_{0}^{z} k(a) G_0(a)\, da \right),
\end{align}\end{subequations}
 where the Wronskian is defined as in Eq.~(\ref{wronskian}).

\begin{figure}
        \centering
                \includegraphics[scale=.5]{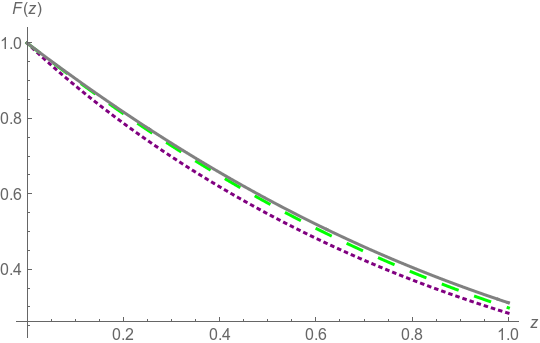}
        \caption{(color online) 
 Exact (solid, gray), perturbation 
(long dashed, green), and 
WKB (dotted, purple)
solutions of $F(z)$ for $\alpha = 1$, $\kappa = 0.7$.
 The perturbation solution closely
 matches the exact solution.} \label{swo2}
\end{figure}

\begin{figure}
        \centering
                \includegraphics[scale=.5]{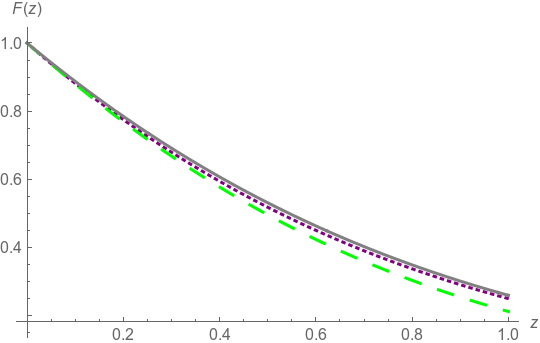}
        \caption{(color online)
 Exact (solid, gray), perturbation 
(long dashed, green), and 
WKB (dotted, purple)
 solutions of $F(z)$ for $\alpha = 1$, $\kappa = 1$. 
The WKB solution closely matches the exact solution.} 
\label{swo3}\end{figure}

\begin{figure}
        \centering
                \includegraphics[scale=.5]{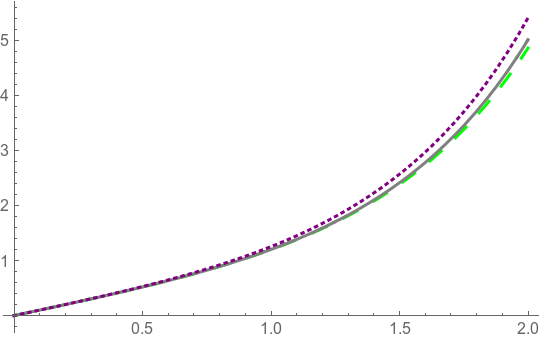}
        \caption{(color online) Exact (solid, gray),  
perturbation (long dashed, green), and 
WKB (dotted, purple) solutions of $G(z)$ for 
$\alpha = 1$, 
$\kappa = 0.8$. 
The perturbation solution closely matches the exact 
solution.} \label{swo4}
\end{figure}

\begin{figure}
        \centering
                \includegraphics[scale=.5]{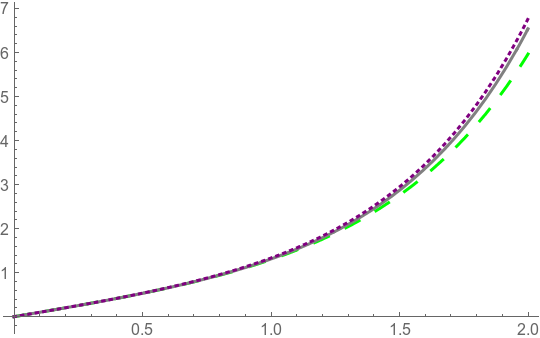}
        \caption{(color online) Exact (solid, gray), perturbation 
(long dashed, green), and 
WKB (dotted, purple) solutions of $G(z)$ for 
$\alpha = 1$, 
$\kappa = 1.1$ The WKB 
solution closely matches the exact solution.} \label{swo5}
\end{figure}

In Figures \ref{swo2}--\ref{swo5} the exact Airy-function
solutions with $\alpha=1$ are compared with the perturbative
and the WKB approximations, for various values of $\kappa$.
As expected, as $\kappa$ grows there is a transition from 
perturbative to WKB regime.
The WKB formulas used here are the first-order ones,
\begin{subequations}\label{WKB1}\begin{align}
F(z) &\approx c_F (\kappa^2 + z^{\alpha})^{-\frac{1}{4}} 
\exp\left[- \int dz \left( \sqrt{\kappa^2 + z^{\alpha}} 
\right) \right], \\
G(z) &\approx c_G (\kappa^2 + z^{\alpha})^{-\frac{1}{4}} 
\sinh\left[ \int dz \left( \sqrt{\kappa^2 + z^{\alpha}} 
\right) \right].
\end{align}\end{subequations}
\goodbreak

\goodbreak

Calculations with special functions determine that
\begin{equation}
W(i,k)=-\,\frac32\,,\quad
c_1= \frac{2b^b}{\Gamma(b)}\,,\quad
c_2 = \frac{\Gamma(b)}{3b^b}\,.
\end{equation}
and hence that
\begin{equation} \label{F'0}
F_0'(0)= -b^{2b-1}\frac{\Gamma(1-b)}
{\Gamma(b)}\,,
\qquad
F_1'(0)=-\,\frac{\sqrt{\pi} b^{1-2b}}{2^{4b-1}}
\, \frac{\Gamma(2b)\Gamma(3b)}{\Gamma(b)\Gamma\left(
\frac12+2b\right)}\,.
\end{equation}
We remark that higher-order $z$-derivatives at $0$ may not 
exist,  because fractional powers of $z$ arise in the
expansion of $k(z)$.

The perturbative expansion of $F_\kappa$ may be
applied with negative values of $\kappa^2$ to
approximate the eigenfunctions, 
solutions of Eq.~(\ref{eigeneq}).

Conceptually it is easy to extend the perturbative
expansions to higher order in $\kappa^2$ by formulas
of the structure
\begin{equation}
F_n(z) = \int dz_1 \cdots dz_n \,
g(z,z_1)\cdots g(z_{n-1},z_n) F_0(z_n).
\end{equation}
We have not implemented any order above the first,
because the iterated numerical integrations would
become  demanding.
We tried a second-order WKB approximation 
 but found, as usual
for a nonconvergent asymptotic series, that it made
the results worse in the region of moderate $z$
where an improvement was most needed.

\section{Outside the wall} \label{sec:outside}

When $z\le0$, ${\cal F}_\kappa(z)$ has the form 
$c_-(\kappa)(e^{-\kappa 
z} + \gamma_-(\kappa)e^{\kappa z})$.
A calculation from the boundary data (\ref{Fdef}) yields
\begin{equation} \label{gammaminus}
\gamma_-(\kappa) = 
\frac{\kappa {\cal F}_\kappa(0) + {\cal F}'_\kappa(0)}
{\kappa {\cal F}_\kappa(0) - {\cal F}'_\kappa(0)}
=\frac{\kappa  + {F}'_\kappa(0)}
{\kappa  - {F}'_\kappa(0)} \,.
\end{equation}
Thus, when both $z$ arguments are negative, 
the Green function (\ref{greenH}) reduces to \cite{HSW} 
\begin{equation} \label{goutside}
g_\kappa(z,z') = \frac{1}{2 \kappa} e^{- \kappa |z-z'|} + 
\frac{1}{2 \kappa}
 e^{\kappa (z+z')} \gamma_{-}(\kappa). 
\end{equation}

 The first 
term in Eq.~(\ref{goutside}) is the Green function for empty 
Minkowski space, which is the same regardless of the presence
 of the wall. 
This term has been exhaustively studied already; more can 
be read about it in almost any introductory text on the Casimir 
effect (for example, \cite{CasReview}) and, with special 
reference to point-splitting regularization, in 
\cite{christensen}. 
Christensen's formula \cite[Eq.~(6.2)]{christensen} does exhibit 
a dependence on the direction of point-splitting in a way that 
differs from one component of $\ev{T_{\mu\nu}}$ to the next and
hence would be considered ``anomalous'' in the sense of 
\cite{Estrada-Mera} if the point separation were to be regarded 
as a physical regularization to be kept nonzero at the end.
That is, however, never done; the direction-dependent terms are 
argued away as artifacts of a noncovariant regularization 
procedure, and the entire Minkowski-vacuum stress tensor is set 
equal to either zero or a multiple of $g_{\mu\nu}$ (cosmological 
constant term) \cite{zeldovich, christensen2, ALN2,  Benasque}.

Returning to the 
second term in Eq.~(\ref{goutside}), one sees that
it contains all of the effects of the wall on space outside the 
wall (i.e., for $z<0$).
Furthermore, because of the rapid decrease of that term
as $\kappa\to\infty$, its contributions to the energy density 
and pressure are continuous and convergent as $\vec\delta\to0$.
So, we may now set $\delta=0$ and $z'=z$, and as pointed out in 
Sec.\ \ref{sec:stress} we may be confident that the two pressures 
parallel to the wall are equal to $-u$. 
The perpendicular pressure $p_z$ now also works out correctly:
In the limit $\delta \rightarrow 0$, the integrand of 
Eq.~(\ref{uandpz})  becomes identically zero for all values of 
$\kappa$, so $p_z = 0$. 
This  confirms
  the principle of virtual work, because the 
total energy
 of the system does not change when the soft wall moves
perpendicularly to itself:  the boundary energy density is 
concentrated 
near the boundary and is merely dragged along with the wall.
(However, the pressure acting on the boundary from the right
from inside the wall remains to be investigated.)
Finally, as in \cite{HSW},  Eq.~(\ref{u}) reduces to
\begin{equation}  \label{uoutside}
u(z) = \frac{1 - 6 \xi}{6 \pi^2} 
\int_{0}^{\infty} d\kappa \kappa^3 e^{2 \kappa z} 
\gamma_{-}(\kappa). 
\end{equation} 
(We emphasize that the Minkowski zero-point energy has 
already been removed from this quantity, which was called
$u(z)-u_0$ in \cite{HSW}.)

The next task is to find a good approximation to 
$\gamma_-(\kappa)$.
 At small $\kappa$ we can use the perturbative 
formulas (\ref{FGpert}) and (\ref{F'0})  to get 
and evaluate
\begin{equation} \label{gam-small}
\gamma_{-}(\kappa) \approx \frac{\kappa + F'_0(0) + \kappa^2 
F'_1(0)}{\kappa - F'_0(0) - \kappa^2 F'_1(0)} .
\end{equation} 
At large $\kappa$, using a WKB formula one 
finds an accidental cancellation in (\ref{gammaminus}) such that
it predicts $\gamma_-$ identically zero. To get a nontrivial 
result, we perform the reweighting of terms carried out in 
  \cite[Sec.~IV]{HSW} to arrive at \cite[Eq.~(4.29)]{HSW}
\begin{equation} \label{gam-large}
\gamma_{-}(\kappa) \approx - \,
\frac{\Gamma(\alpha + 1)}{(2 \kappa)^{\alpha + 2}}. 
\end{equation}
(It is clear that this leading term would appear in a WKB 
calculation only at a rather high order, increasing 
with~$\alpha$.)

For $\alpha=1$ and $2$
these two approximations can be tested against exact 
calculations 
from Eqs.\ (\ref{airyF}) and (\ref{parcylF}).  
They  match  $\gamma_{-}(\kappa)$ 
very well for small and large  $\kappa$, respectively, 
but there 
is a significant intermediate region in which neither is 
accurate. After some experimentation, we remedied this very 
well by introducing the  spline function 
\begin{equation} s(\kappa) = e^{a + b \kappa}, 
\label{spline}\end{equation} 
\noindent where $a$ and $b$ are  chosen so that
the functions match at the endpoints of a central interval
$[d_1,d_2]$, 
\begin{align} s(d_1) &= 
\frac{ F'_0(0) +d_1 + d_1^2 
F'_1(0)}{F'_0(0) -d_1 + d_1^2 F'_1(0)}  
  \\ s(d_2) &=  
\frac{\Gamma(\alpha + 1)}{(2 d_2)^{\alpha + 2}}\,, \end{align} 
 \begin{figure} 
	\centering
	\includegraphics[scale=.5]{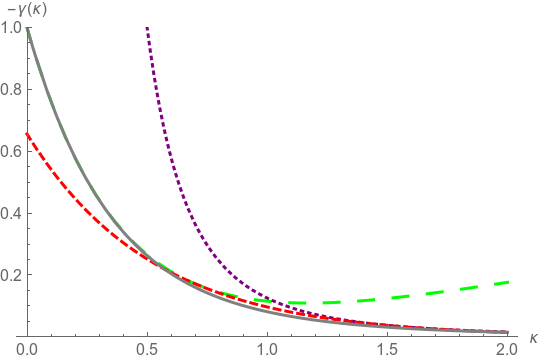}
	\caption{(color online) Approximations (perturbation in 
long green dashes, Milton (\ref{gam-large}) 
 in purple dots, spline in red dashes) and exact 
solution 
(solid gray) for $-\gamma_-$ in the case $\alpha = 1$.} 
\label{swo6}\end{figure}
\begin{figure}
	\centering
	\includegraphics[scale=.5]{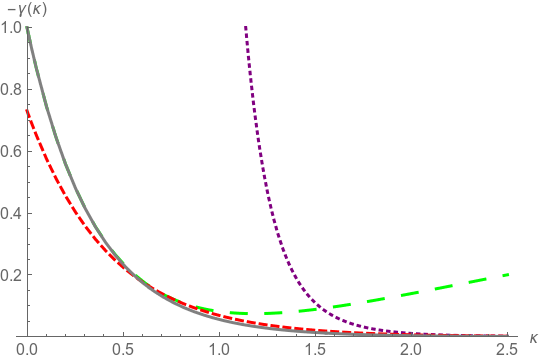}
	\caption{(color online) Approximations 
(perturbation in long green dashes, Milton
 in purple dots, 
spline in red dashes) for
$-\gamma_-$ in the case $\alpha = 6$. 
A numerical approximation of the exact solution
is shown in solid gray. Notice that the spline 
 matches the exact function well even  when the two 
asymptotic approximations are far apart.}
 \label{swo7}\end{figure}
and then $d_1$ and $d_2$ are chosen so that the derivatives also 
match there (by an application of Newton's method in 
\emph{Mathematica}).
(The exponential spline is superior to a linear one because it 
qualitatively reproduces the obvious convex behavior of 
$\gamma_-(\kappa)$.)
We approximate $|\gamma_{-}(\kappa)|$ by a 
piecewise-defined function given by the perturbation solution 
when 
$\kappa < d_1\,$, the spline function (\ref{spline})
when $d_1 < \kappa < d_2\,$, 
and the large-$\kappa$ solution when $d_2 < \kappa$. 
Since this piecewise function is seen to be strictly greater
 than the exact solution, our value for $u(z)$ will be an 
 upper bound. 
 Some plots of these approximations are 
given in Figs.\ \ref{swo6} and~\ref{swo7},
with comparisons to the exact Airy formula for $\alpha=1$
and to a numerical solution for $\alpha=6$.

\begin{figure}[h!] 
	\centering
	\includegraphics[scale=.5]{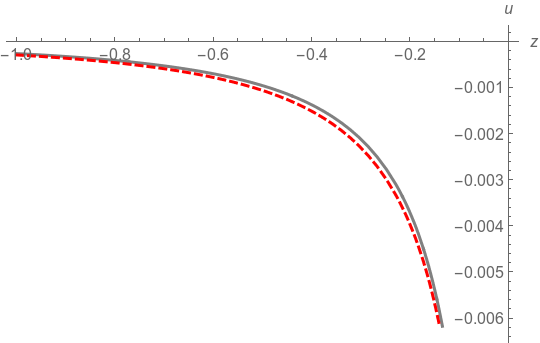}
	\caption{(color online) Approximation (dashed) and exact 
solution (solid)
 for $\ev{T_{00}}$ outside the wall in the $\alpha = 1$ case. 
The factor $1 - 6\xi$ is omitted.
The vertical axis is an asymptote.}
\label{swo8}\end{figure}

\begin{figure}[h!]  
	\centering
	\includegraphics[scale=.5]{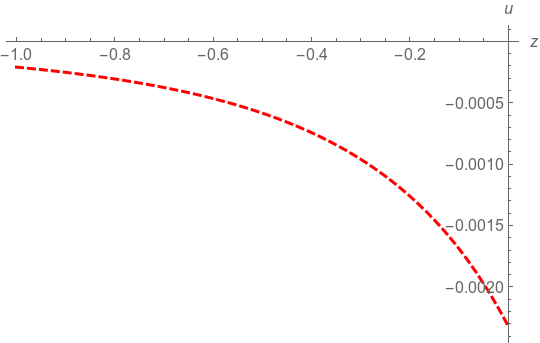}
	\caption{(color online) Approximation for $\ev{T_{00}}$ 
outside the wall 
in the 
$\alpha = 6$ case. The factor  $1-6\xi$ is omitted.
The vertical axis has an intercept.}
 \label{swo9}\end{figure}

Now that we have  $\gamma_{-}(\kappa)$, we can 
find the energy density outside the wall by  numerical 
integration of Eq.~(\ref{uoutside}).
Representative resulting plots are in Figs.\  \ref{swo8}
and~\ref{swo9}.
As anticipated in \cite{HSW}, the integral for 
 $u(0)$  converges only for $\alpha > 2$.
The present method is a significant improvement over the 
preliminary treatment in \cite{HSW} in regards to the 
contributions from small~$\kappa$.

\section{Inside the wall} \label{sec:inside}

The problem of finding the energy density inside the wall, 
where the 
potential is nonzero, is substantially more difficult than the 
case outside the wall. First, we need to compute not only
the coefficient $\gamma_+(\kappa)$ but also the solutions
$F_\kappa(z)$ and $G_\kappa(z)$, functions of two variables.
Second, there are divergences ``in the bulk'' proportional
to local functionals of the potential,
which need to be identified and removed by renormalization
in the usual sense of quantum field theory \cite{MNS}.
Third, the Green function does not divide neatly and
uniquely into two terms, one purely divergent and the
other containing all the real physics, as happened in
Eq.~(\ref{goutside}).
Finally, at small $\kappa$ it is difficult to normalize
WKB solutions properly, since the boundary conditions
(\ref{Fdef}) and (\ref{Gdef}) must be applied at $0$,
precisely where the WKB approximation is least valid.
It is this last nuisance that forces us to consider the
whole family of more general normalization conventions.

We begin, as we did in the previous section, by expressing the
Green function more explicitly in terms of the basis functions.
Let
\begin{equation}
H_\kappa = c_+(\kappa)({\cal G}_\kappa + 
\gamma_+(\kappa){\cal F}_\kappa)
\end{equation}
(thereby defining $c_+$ and $\gamma_+$).
  Then
\begin{equation}\label{gammaplus}
\gamma_+(\kappa)=-\,
\frac{\kappa {\cal G}_\kappa(0) -{\cal G}_\kappa'(0)}
{\kappa {\cal F}_\kappa(0) -{\cal F}'_\kappa(0)}
=\frac1{\kappa-F'_\kappa(0)}\,.
\end{equation}
It follows that
\begin{equation}\label{ginside1}
	g_{\kappa}(z,z') 
= c_+(\kappa)\frac{\left({\cal G}_\kappa(z_<) + 
\gamma_{+}(\kappa) {\cal 
F}_\kappa(z_<)\right){\cal F}_\kappa(z_>)}
{W({\cal F}_\kappa,e^{\kappa z})}
= \left(G_\kappa(z_<) + \gamma_{+}(\kappa) F_\kappa(z_<)
\right) F_\kappa(z_>).
\end{equation}
Because $W({\cal F},e^{\kappa z}) = c_+ W({\cal F},{\cal G})$,
the middle (generic) version of Eq.\ (\ref{ginside1})
reduces to
\begin{equation}\label{ginside2}
	g_{\kappa}(z,z') 
= \frac{{\cal G}_\kappa(z_<){\cal F}_\kappa(z_>)}
{W({\cal F}_\kappa,{\cal G}_\kappa)}
 + \gamma_{+}(\kappa) \frac{{\cal F}_\kappa(z)
 {\cal F}_\kappa(z')}
{W({\cal F}_\kappa,{\cal G}_\kappa)},
\end{equation}
which is \cite[Eq.~(4.5)]{HSW}.
Here the generic formula for $\gamma_+$
(the middle member of Eq.\ (\ref{gammaplus})) must be used.

Because of the freedom to redefine $\cal G$ by adding a
multiple of $\cal F$, the division of $g_\kappa$ in 
Eq.~(\ref{ginside2}) into two terms is rather arbitrary.
In the extreme case that ${\cal G} = {\cal H}$,
the second term vanishes completely.
Ideally one would like to go to the opposite extreme and 
make the first term ``purely divergent''\negthinspace,
so that all the interesting physics is in the second term.
(Note that, because of rapid decay as function of $\kappa$,
the second term yields convergent integrals that are 
continuous functions of the point separation $\vec\delta$,
so one can set the points equal before trying to 
evaluate them.)
The convention ${\cal G}_\kappa(0)=0$ is certainly not that 
ideal; it makes the first term in $g_\kappa$ into the
Green function for a problem with a Dirichlet
condition at $z=0$, and the resulting vacuum stress will
have the singular behavior associated with a perfectly 
reflecting boundary there.
Since our real problem does not have such a boundary,
but only a mild coefficient singularity that weakens
with increasing $\alpha$ 
(see \cite[p.~25]{WKBwall} and \cite[p.~146]{SpecGeom}), 
these boundary terms must be
cancelled by contributions from the second term.

We present here a preliminary numerical exploration in the
spirit of the previous section, 
 starting  with lowest-order approximations for 
$\gamma_+(\kappa)$.
 One can use the WKB form of $F(z)$ in the high $\kappa$ 
region to find the leading term
\begin{equation}
	\gamma_{+}(\kappa) \approx \frac{1}{2 \kappa}\,.
\end{equation}
In the region of small $\kappa$, one approximates $F_\kappa'(0) 
\approx F_0'(0)$ to get
\begin{equation}
	\gamma_{+}(\kappa) \approx \frac{1}{\kappa + b^{2b - 1}
 \Gamma(1 - b) / \Gamma(b)}
\end{equation}
where
$	b = (\alpha + 2)^{-1}$ as before.
\begin{figure}
	\centering
	\includegraphics[scale=.8]{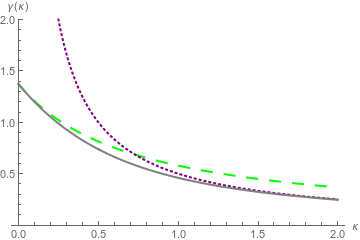}
	\caption{(color online) $\gamma^{+} (\kappa)$ for 
$\alpha=1$. The solid gray line is the 
exact solution, the long-dashed green line is the 
perturbation expansion, and 
the dotted purple line is the WKB approximation.}
\label{swo10}\end{figure}
Although the results in Fig.\ \ref{swo10} are as expected,
we then found that the spline technique we used for $\gamma_-$
does not work as well for $\gamma_+\,$;  details would
be premature here.

\begin{figure}
	\centering
	\includegraphics[scale=.8]{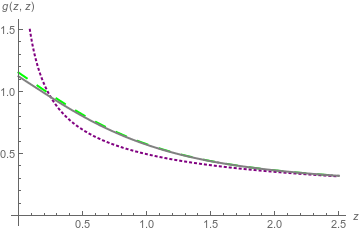}
	\caption{(color online) $g_{\kappa} (z,z)$ for $\alpha=1$ 
and $\kappa = 0.15$. 
The solid gray line is the 
exact solution, the long-dashed green line is the 
perturbation expansion, and 
the dotted purple line is the WKB approximation.
The perturbation expansion is quite good here throughout our 
range of $z$, whereas the WKB expansion takes a long time to 
converge to the correct value.}
\label{swo11}\end{figure}

\begin{figure}
	\centering
	\includegraphics[scale=.8]{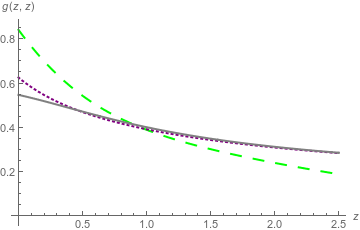}
	\caption{(color online) $g_{\kappa} (z,z)$ for $\alpha=1$ 
and $\kappa = 0.8$. 
The solid gray line is the 
exact solution, the long-dashed green line is the 
perturbation expansion, and 
the dotted purple line is the WKB approximation.
The perturbation expansion is very poor here, 
but the WKB expansion quickly 
converges to a reasonable value.}
\label{swo12}\end{figure}

We proceed to tackle the Green function.
In creating Figs.\ \ref{swo11}--\ref{swo12}
we used the perturbative form of $\gamma_{+}(\kappa)$ 
and the perturbative solutions 
(\ref{FGpert}), (\ref{pert1}) for $F_\kappa(z)$ and $G_\kappa(z)$ 
to obtain a ``perturbative
approximation''\negthinspace, 
and the WKB form of $\gamma_{+}(\kappa)$ and the WKB solutions 
(\ref{WKB1}) for $F_\kappa(z)$ and $G_\kappa(z)$ to obtain a 
``WKB approximation''\negthinspace.
As expected, Fig.\ \ref{swo11} shows that the perturbative
expansion is good for a small~$\kappa$,  and 
Fig.\ \ref{swo12} shows that the WKB expansion
is good for a large~$\kappa$.
Ultimately, there is no reason why 
the approximation used for $F(z)$ and $G(z)$ must be the
same as that used for $\gamma_+\,$, which implicitly
involves $z=0$, since the location of the transition to
WKB behavior depends on~$z$.

\goodbreak
\section{Conclusion} \label{sec:concl}

The power wall is a promising model of the effects of 
a boundary on the vacuum state of a quantum field.
It helps to separate the genuine effects from the pathologies
associated with idealized boundary conditions.

This paper deals with a flat wall interacting with a scalar
 field.
We have set up a systematic notation for the solutions
of the separated field equation and  developed useful
approximations for them in the limits of large and small
imaginary wavenumber $\kappa$, 
which have been tested numerically.
In the region outside the wall, we obtained approximations
for the energy density that closely match the exact solutions
in both asymptotic limits and are reasonably accurate
in between.  

Progress in the region inside the wall has been slow,
because of both technical problems and the conceptual 
complication of renormalization in the presence of a nontrivial
potential.  The renormalization problem has been investigated
separately \cite{MNS,self} for a general potential;
\cite{self} uses
higher-order WKB approximations, with special attention to
the application to the power wall.  Precise calculation of the 
renormalized stress tensor, including contributions from the
non-WKB region of the spectrum, are in process.

The foremost motivation for this work was to resolve the
``pressure anomaly'' in the force on a perpendicular  wall
that has been observed \cite{Estrada-Mera} 
in calculations for a hard wall regularized by a finite cutoff.
Outside the soft wall, we have calculated the energy and 
pressure and shown that the anomaly does not occur in this 
model.  (Eventual success inside the wall is expected.)
We attribute this success to the fact that the soft-wall 
model is 
a consistent physical system whose energy density is 
well behaved from the start, 
rather than being forced to be finite by
 an \emph{ad hoc} cutoff.

Calculations so far (here and in \cite{HSW,thesis,self}) deal 
primarily with small values of $\alpha$ and hence with the 
extraneous divergences near $z=0$ caused by the singularity in 
the potential there.  For very large values of $\alpha$ these 
divergences must disappear, and the stress tensor must  
resemble that of a hard wall near $z=1$. Until such computations 
are available, it is not meaningful to make a comparison of our 
results with others, such as those of Barton 
or Passante et al.\ (see Appendix).

 \bigskip
 \begin{acknowledgments} 
 This research was supported by National Science Foundation 
Grant   PHY-0968269 and 
undergraduate research funds from the Texas A\&M University 
Department of Physics and Astronomy.
The participation of Jef Wagner was made possible by a 
temporary Visiting Assistant Professorship 
in the TAMU Department of Mathematics. 
Valuable comments by  Lev Kaplan, Kim Milton, and 
Prachi Parashar are gratefully acknowledged.

\end{acknowledgments}

\appendix

 \section{Historical perspective}

First to base the analysis of the Casimir effect on local 
quantities (i.e., the expectation values of components of the 
stress tensor) rather than total energy or force were Brown and 
Maclay \cite{BmC}.  DeWitt \cite{dewitt} recognized the 
importance of this topic for understanding quantum field theory 
in curved space-time.  Consequently, the generic behavior of the 
stress tensor near boundaries was investigated from differing 
points of view by two groups in Austin \cite{DC,KCD}; Deutsch and 
Candelas \cite{DC} argued that divergences in the total energy 
are related to finite and physically meaningful, though 
nonintegrable, distributions of energy density near the boundary 
and, therefore, must be taken seriously.  This can be regarded as
the origin of the modern program of softening (not discarding) 
surface terms by improved modeling of the boundaries themselves.
Later, the analysis of the energy in a box in terms of classical 
paths by Hertzberg et al.\ \cite{HJKS} demonstrated that energy 
density near boundaries has physical meaning and can be 
responsible for counterintuitive signs in some parts
of Casimir forces.

One of the most ambitious programs in the direction of 
improved modeling has been conducted by Barton 
\cite{barton01,barton04a,barton04b,BartonB,barton05b}, 
studying nonlinear 
interactions of the field with degrees of freedom inside the 
walls.  

Closer to the spirit of the present work is 
Ford and Svaiter's  investigation of a hard boundary 
whose location is subjected to random fluctuations  \cite{FS}.
Similar work has been done recently by Passante et al.\
\cite{pas1,pas2,pas3}

Qualitatively similar results appear in a
large number of papers by Graham, Jaffe, Olum,  
and coworkers (such as Refs.~\cite{GJKQSW,OG,GO,Jaffe})
 culminating in the book \cite{GQW}.
That program differs from ours superficially by dealing 
with 
a high, narrow potential hill (instead of a one-sided wall), and 
more fundamentally in the choice of calculational methods
 (techniques of scattering theory instead of local, 
 differential-equation analyses), which results in a rather 
different point of view.
An earlier  paper in that tradition  is  Bordag \cite{bordag}.

The present program is rooted in \cite{rect}, which proposed to 
model a soft wall simply by maintaining a finite ultraviolet 
cutoff in calculations for a hard wall.  Results were 
qualitatively so similar to those of Ford and Svaiter \cite{FS} 
and Graham and Olum \cite{OG,GO} as to suggest that all the 
approaches were on the same track.  However, in calculations for 
a spherical boundary, Martin Schaden and Fulling (unpublished, 
but summarized in \cite{norman09}) discovered a pressure 
anomaly --- a violation of Eq.\ (\ref{balance}).
Detailed examination \cite{Estrada-Mera, Benasque} showed that 
this problem existed for flat boundaries also, and that it was 
related to the direction dependence of vacuum energy as 
regularized by point-splitting \cite{christensen}. 
That is the motivation for the study of the soft wall 
\cite{SpecGeom,HSW,Benasque,thesis,self}.

 The closest work to ours that we know of in previous literature 
(until \cite{HSW} and \cite{MNS}) is that of
Actor and Bender \cite{AB}, in which the perfectly reflecting 
wall is replaced by a harmonic-oscillator potential.
 That paper was written before the modern critiques of formal 
renormalization \cite{BartonA,Jaffe} and the modern emphasis on 
local quantities (such as energy density).  It deals with total 
energies calculated by zeta-function regularization.

Here we have not reviewed papers on scalar quantum field theory 
in general background scalar potentials and the resulting issues 
of renormalization, since that topic relates more to \cite{self} 
and possible later papers.
Nor have we listed the large body of papers on fields interacting 
with slabs of material (such as dielectrics) or delta-function 
potentials; such models are not truly ``soft'' by our definition, 
since boundary divergences remain.
The remarks on the works of Graham et al.\ and Actor and Bender 
are paraphrased from \cite{Benasque}.

 \goodbreak

 \end{document}